\documentclass[runningheads]{llncs}

\pdfoutput=1
\usepackage[pdftex]{graphicx}
\usepackage{amsmath}

\usepackage{algorithm,algpseudocode}
\algnewcommand\algorithmicinput{\textbf{Input:}}
\algnewcommand\algorithmicoutput{\textbf{Output:}}
\algnewcommand\Input{\item[\algorithmicinput]}%
\algnewcommand\Output{\item[\algorithmicoutput]}%

\begin{document}

\title{Graph Drawing with Morphing Partial Edges}

\author{Kazuo Misue \and
Katsuya Akasaka}
\authorrunning{K. Misue and K. Akasaka}

\institute{University of Tsukuba, Tsukuba, Japan\\
\email{misue@cs.tsukuba.ac.jp, akasaka@vislab.cs.tsukuba.ac.jp}}
\maketitle              
\begin{abstract}
A partial edge drawing (PED) of a graph is a variation of a node-link diagram.
PED draws a link, which is a partial visual representation of an edge,  and reduces visual clutter of the node-link diagram. 
However, more time is required to read a PED to infer undrawn parts.
The authors propose a morphing edge drawing (MED), which is a PED that changes with time. 
In MED, links morph between partial and complete drawings; thus, a reduced load for estimation of undrawn parts in a PED is expected. 
Herein, a formalization of MED is shown based on a formalization of PED.
Then, requirements for the scheduling of morphing are specified.
The requirements inhibit morphing from crossing and shorten the overall time for morphing the edges. 
Moreover, 
an algorithm for a scheduling method implemented by the authors is illustrated and the effectiveness of PED from a reading time viewpoint is shown through an experimental evaluation.
\end{abstract}

\section{Introduction}

The partial edge drawing (PED) of a graph is a variation of a node-link diagram that is a visual representation of a graph.
In PED, a link is drawn, which is a partial visual representation of an edge; that is, a part of the link is omitted, and then intersections of links are eliminated. 
Therefore, PED can reduce visual clutter of node-link diagrams.
An experimental evaluation by Bruckdorfer et al. shows that PEDs reduces errors and provides higher accuracy when reading graphs than traditional node-link diagrams; however, longer reading time is required~\cite{bruckdorfer2015ped}. 

We propose a morphing edge drawing (MED), which is a PED that changes with time.
In MED, links are morphed between partial and complete drawings.
Therefore, reduced loads are expected to infer undrawn parts in a PED. 
However, the effect depends on the scheduling of morphing edges.
We designed a scheduling algorithm that did not unnecessarily cause links to cross. 
Then, we performed a user study to evaluate the effectiveness of MEDs by implementation.
The contributions herein are as follows: 
Proposal and formalization of MED. 
Setting scheduling requirements. 
Proposal of algorithm for scheduling morphing. 
Evaluation of MED via user study.

\section{Partial Edge Drawing}\label{sec:ped}

Let $G=(V,E)$ be a simple undirected graph and let $\Gamma(G)=(\Gamma(V),\Gamma(
E))$ be a drawing of $G$,
where $\Gamma(V)=\{\Gamma(v)|v\in V\}$ and $\Gamma(E)=\{\Gamma(e)|e\in E\}$.
Let
$\Gamma(G)$ be a traditional straight-line drawing.
Let the drawing $\Gamma(v)$ of a node $v\in V$ be a small disk placed at a position $p_{v}$ and
let $\Gamma(e)$ of an edge $e\in E$ be a straight-line segment between two nodes (disks) incident to the edge.
That is, $\Gamma(e)=\{s\cdot p_{w}+(1-s)\cdot p_{v}|s\in [0,1]\}$ when $e=(v,w)$.
We call $\Gamma(G)$ a {\em complete edge drawing (CED)} because it draws every straight-line representing an edge completely.

We express the partial drawing of an edge $e=(v,w)$ as a function $\gamma_{e}: [0,1]^2 \rightarrow 2^{\Gamma(e)}$ shown in Exp.~(\ref{eq:gammae}).
\begin{equation}
  \gamma_{e}(\alpha,\beta) =
  \begin{cases}
    \{s\cdot p_{w}+(1-s)\cdot p_{v}|s\in [0,\alpha]\cup [\beta,1]\} & \text{for $\alpha < \beta$} \\    
    \Gamma(e)    & \text{for $\alpha \ge \beta$}
  \end{cases}  
\label{eq:gammae}
\end{equation}
That is, $\gamma_{e}(\alpha,\beta)$ of edge $e$ comprise the parts that remain after removing the corresponding parts $(\alpha,\beta)$ from $\Gamma(e)$ when the entire $\Gamma(e)$ corresponds to the interval $[0,1]$.
Each of the remaining continuous parts is called a {\em stub}.
The parameters $\alpha$ and $\beta$, which determine the stub lengths, are {\em partial edge parameters}.
When $0<\alpha$ and $\beta<1$ for $\gamma_{e}$ , the part to be deleted is not the end of $\Gamma(e)$; two stubs remain at the two nodes incident to the edge $e$. 
These are called a {\em pair of stubs}.

Drawing $\Gamma_{PED}(G)=(\Gamma(V),\Gamma_{PED}(E))$ is a {\em partial edge drawing (PED)} if for all edges $e\in E$, $\alpha_{e}$ and $\beta_{e}$ are given, and at least an edge $e_{1}\in E$ exists with $\alpha_{e_{1}}<\beta_{e_{1}}$,
where $\Gamma_{PED}(E)=\{\gamma_{e}(\alpha_{e},\beta_{e})|e\in E\}$.
When $\alpha_{e} = 1-\beta_{e}$, i.e., the lengths of a pair of stubs are the same, the drawing is a {\em symmetric PED (SPED)}.
The smaller parameter $\alpha_{e}$ is the {\em stub-edge ratio}.
If the stub-ratios for all edges are the same $\delta$,
the drawing is a {\em $\delta$-symmetric homogeneous PED ($\delta$-SHPED)}.

Herein, we assume that $\Gamma(G)$ is given in advance and stubs may have intersections.

\section{Related Work}

Becker et al. conceived a drawing concept in which only half the links are used to reduce the visual clutter during the development of a tool called SeeNet~\cite{becker1995visualizing}.
Parallel Tagcloud, developed by Collins et al., adopts a method similar to PED~\cite{collins2009parallel}.
Although Parallel Tagcloud is an extension of Tag Cloud, it can be regarded as a hierarchical layout of directed graphs; thus, it is useful against visual clutter caused by crossing of links.
They can avoid drawing intersections by representing links as straight lines without drawing in the middle of the links.

Bruckdorfer et al. formalized the PED in their study~\cite{bruckdorfer2012mad};
our formalization of PED in Section~\ref{sec:ped} is a modified version of their formalization.
They added continuity of the omitted parts of links as a condition, we have incorporated this into the formalization herein.
Moreover, although they focused on a layout without crossing stubs in PED, this study allows stub crossings.
Burch et al. applied PED to directed graphs using tapered links~\cite{burch2012evaluating}.
Schmauder et al. applied PED to weighted graphs by representing weights with edge colors~\cite{schmauder2015visualizing}.

Bruckdorfer et al. performed a comparison between CED\footnote{They call it the traditional straight-line model (TRA).} and 1/4-SHPED on reading performance of graphs~\cite{bruckdorfer2015ped}.
Although the statistical significance was not shown, from their chart that visualizes the experimental results, in the task of reading graphs (for adjacency check of two nodes or search for adjacent nodes), we can guess that 1/4-SHPED is slightly more accurate than CED; however, the response time of 1/4-SHPED is longer.
Binucci et al. conducted a more detailed evaluation to reveal that SHPED has high accuracy in reading graphs within SPED~\cite{binucci2016partial}.
Burch examined the effect of stub orientation and length on graph reading accuracy~\cite{burch2017user}. 

Blass et al. avoided using arrows to facilitate grasping high-dimensional transitions in the state transition diagram and proposed moving the dashed pattern with animation~\cite{blass2009smooth}.
Holten et al. compared the recognition accuracy of graphs in various edge drawing methods, such as tapered links and curved links, including animation~\cite{holten2011extended}.
They showed that the recognition accuracy of the graph is high by representation using animation.
Romat et al. attempted to extend the design space using animation of edge textures~\cite{Romat:2018:AET:3173574.3173761}.
The proposal herein can be considered as an application of animation to graph drawing, especially to drawing edges.
However, the purpose is not to express the orientation, but to improve the reading accuracy and efficiency for graphs.

\section{Morphing Edge Drawing}

Let $T$ be a set of times.
Then, function $\mu_{e}: T \rightarrow 2^{\Gamma(e)}$, which determines a partial drawing of edge $e$ for time $t\in T$, a {\em morphing function}.
A dynamic drawing $\Gamma_{MED}(G)=(\Gamma(V),\Gamma_{MED}(E))$ of graph $G$ with morphing functions is a {\em morphing edge drawing (MED)},
where $\Gamma_{MED}(E)=\{\mu_{e}|e\in E\}$ is a set of morphing functions. 

Then, a function $\rho_{e}: T\rightarrow [0,1]^{2}$, which determines the partial edge parameters for a time $t\in T$, is a {\em ratio function}.
The morphing function $\mu_{e}$ can be constructed as $\mu_{e}(t) = \gamma_{e}(\rho_{e}(t))$ using the ratio function.

\subsection{Symmetric MED}

When all $\rho_{e}$ for all $e\in E$ satisfies $\rho_{e}(t)=(\delta_{t}, 1-\delta_{t})$ ($0\le \delta_{t} \le 1/2$) for all $t\in T$, we get SPED at any time. 
Thus, such a ratio function is a {\em symmetric ratio function};
furthermore, if a MED is composed of symmetric ratio functions, it is referred to as a {\em symmetric MED (SMED)}.
As the two values obtained by a symmetric ratio function depend on each other, we can define the function as $\rho_{e}: T\rightarrow [0,1/2]$ without ambiguity. 

\begin{figure}[htb]
  \centering
  \includegraphics[scale=0.3]{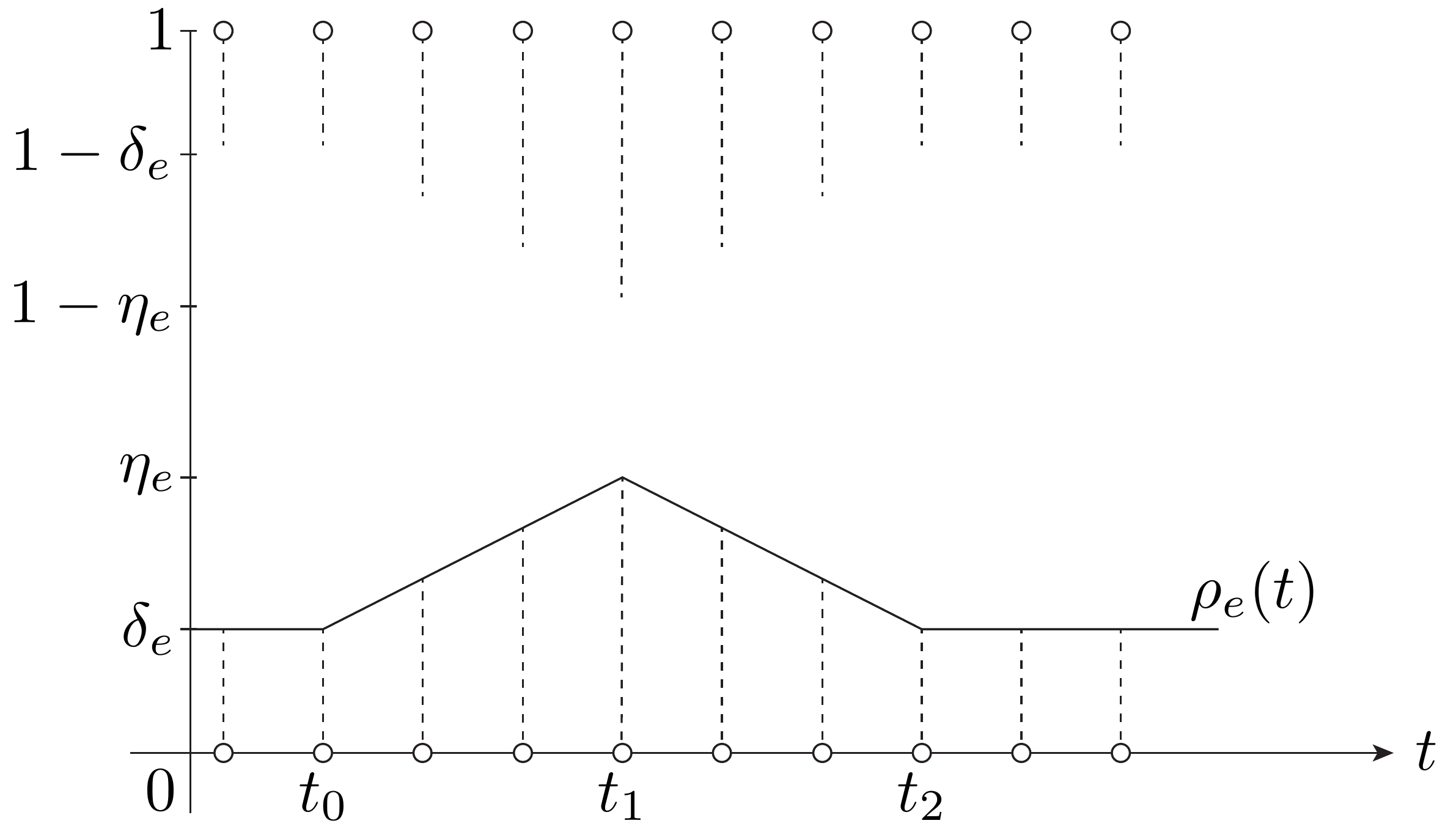}
  \caption{Graph of $\rho_{e}$. Each pair (top and bottom) of dashed lines represent a pair of stubs expanding and contracting.}
  \label{fig:rho_graph}
\end{figure}

Morphing of edge $e$ extending from stub-edge ratio $\delta_{e}$ to $\eta_{e}$ and then shrinking to $\delta_{e}$ is expressed by a symmetric ratio function $\rho_{e}$, expressed as Exp.~(\ref{eq:sigma_e1}), where $t_{0}$ is the start time of the morphing, $l$ is the length of $\Gamma(e)$, and $s$ is the speed of the tips of the stubs ({\em morphing speed}).
Here, let the morphing speed be constant. 
Fig.~\ref{fig:rho_graph} shows the graph of the function. 
\begin{equation}
  \rho_{e}(t) =
  \begin{cases}
    \delta_{e}                & \text{for $t\le t_{0}$ or $t_{2}<t$} \\
    \delta_{e} + (t - t_{0})s/l & \text{for $t_{0}<t\le t_{1}$} \\
    \eta_{e} - (t - t_{1})s/l & \text{for $t_{1}<t\le t_{2}$},
  \end{cases}  
\label{eq:sigma_e1}
\end{equation}
where $t_{1}$ is the time when the stub-edge ratio becomes $\eta_{e}$, and $t_{2}$ is the time when the stub-edge ratio returns to $\delta_{e}$.
Using one-way travel time $d_{1}=(\eta_{e} - \delta_{e})l/s$, $t_{1}=t_{0}+d_{1}$ and $t_{2}=t_{0}+2d_{1}$.

If the ratio functions have the same $\delta$ and $\eta$ for all edges, the drawing is a {\em $(\delta,\eta)$-symmetric homogeneous MED ($(\delta,\eta)$-SHMED)}.
Note, this homogeneity does not always mean synchronicity.
Drawings by $(\delta,\eta)$-SHMED may not be SHPED at any time. 

When $\eta=1/2$, we omit $\eta$, like {\em $\delta$-symmetric homogeneous MED ($\delta$-SHMED)}.
In the drawing by $\delta$-SHMED, a pair of stubs of edge $e$ becomes $\Gamma(e)$ at a certain moment.
Intuitively, the drawing by $\delta$-SHMED changes between $\delta$-SHPED and CED.
However, a moment when it becomes CED does not always exist.

\section{Scheduling of Morphing}

We have set requirements to design the scheduling of morphing of all edges as follows:

\subsubsection{R1: Morphing does not make crossings}
To maintain the reading accuracy, visual clutter should be minimized.
Therefore, morphing should not result in new crossings among stubs.
However, if another edge exists that crosses a stub with a stub-edge ratio $\delta$, then the crossing is inevitable.
As mentioned earlier, rearrangements to avoid such crossings are beyond the scope of this paper.
The requirement is to avoid crossings in the center area undrawn by $\delta$-SHPED, (we refer to these areas as {\em blank areas}).

\subsubsection{R2: Shorten morphing time for all edges}
The time taken for a viewer to focus on a stub should be minimized before morphing of the stub.
We do not know in advance which stubs the viewer will focus on.
Therefore, it is necessary to repeat morphing of all edges, and it is necessary to shorten the total morphing time of all edges.

\subsection{Morphing Group}
\begin{figure}[htb]
  \centering
  \includegraphics[scale=0.33]{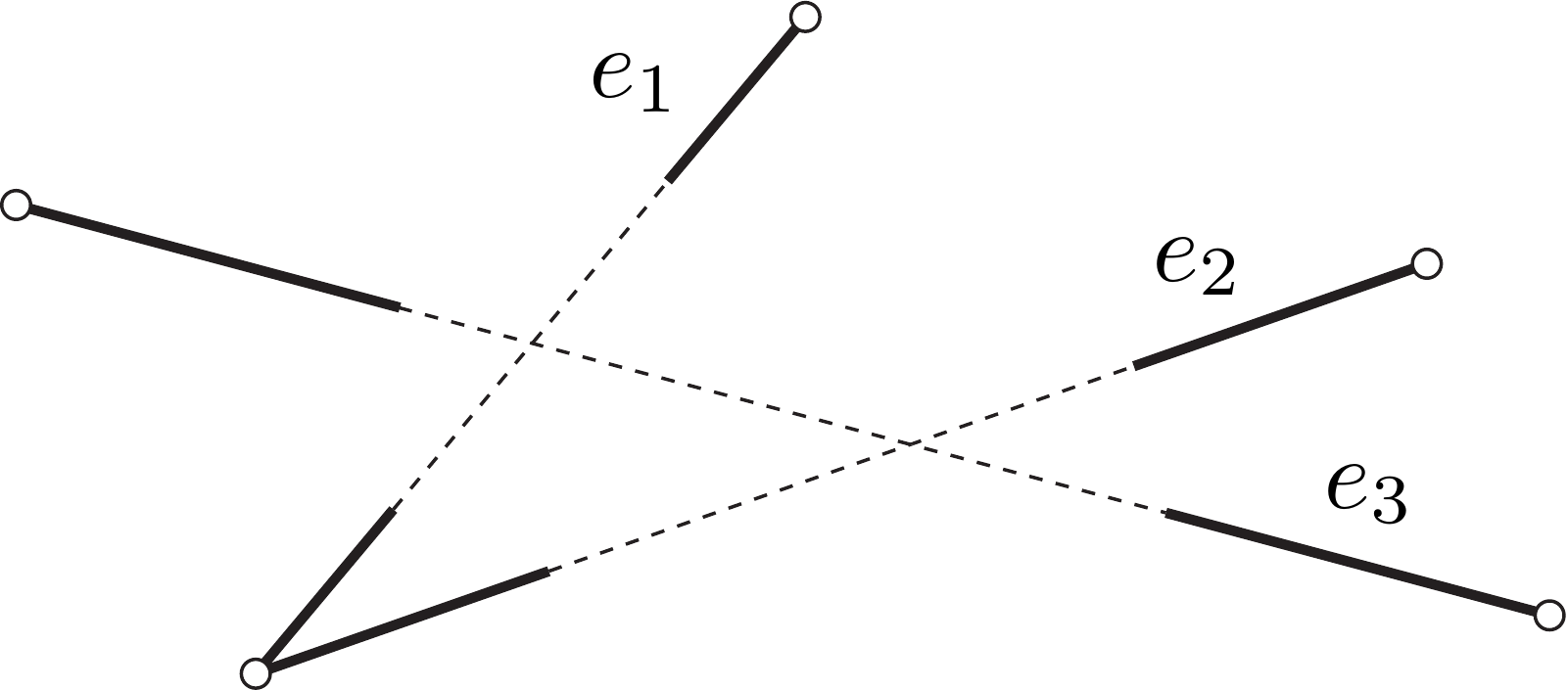}
  \caption{Dependency between edges. The edges $e_{1}$ and $e_{2}$ do not intersect, but both intersect $e_{3}$, so they cannot be morphed independently. Dotted lines represent the omitted parts (blank areas).}
  \label{fig:dependency}
\end{figure}

First, two non-crossing edges do not generate new crossings of stubs at any timing when morphing,
i.e., they can morph simultaneously and independently.
However, two edges that do not intersect may not be able to morph independently, depending on the relationship with other edges (see Fig.~\ref{fig:dependency}).

A set of edges where the timing of morphing may affect each other is a {\em morphing group}.
Different morphing groups can be scheduled independently.
To determine the morphing groups, another graph is generated from the graph to be drawn.
To avoid ambiguity, we will express the newly generated graph and its components as $\langle$graph$\rangle$, $\langle$node$\rangle$, and $\langle$edge$\rangle$.
Let each edge be a $\langle$node$\rangle$.
Suppose that there is an $\langle$edge$\rangle$ between $\langle$nodes$\rangle$ (i.e., edges) that intersect each other in a blank area.
$\langle$Nodes$\rangle$ (i.e., edges) included in each connected component of the $\langle$graph$\rangle$ generated in this manner constitute a morphing group.

As two edges belonging to different morphing groups do not intersect, it is possible to schedule them independently in units of morphing groups.
Hereafter, we describe scheduling of morphing of edges included in a morphing group.

\subsection{Sequential Morphing}

To prevent new stub crossings from being generated by morphing of two edges intersecting in blank areas, entry into a blank area should be exclusive.
That is, the safest scheduling, satisfying requirement R1 (morphing does not result in crossings), is to perform edge morphing sequentially.
However, with such simple scheduling, R2 cannot be satisfied.

\subsection{Packing Morphing Intervals}

\begin{figure}[htb]
  \centering
  \includegraphics[scale=0.33]{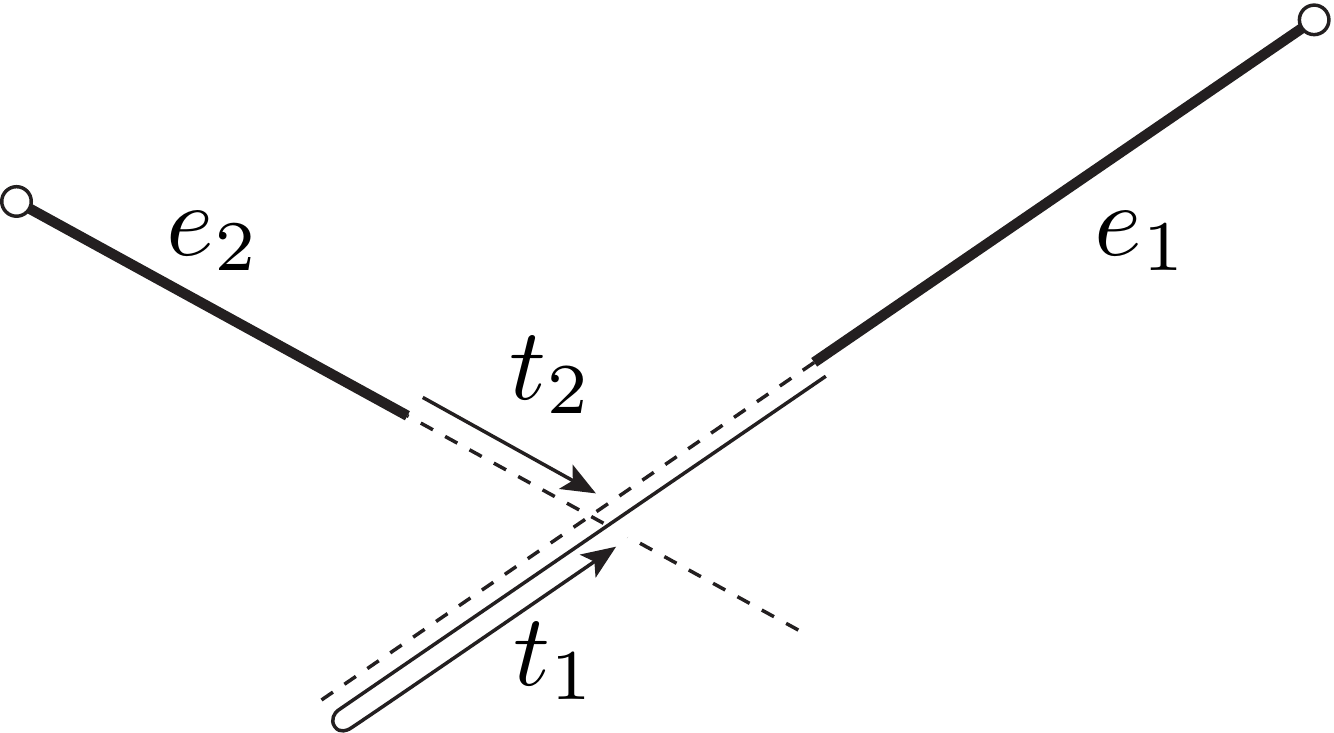}
  \caption{Packing morphing intervals.}
  \label{fig:just_non_crossing}
\end{figure}

Assume there are two intersecting edges $e_{1}$ and $e_{2}$, as shown in Fig.~\ref{fig:just_non_crossing}.
If the scheduling is such that when a stub of edge $e_{1}$ has been stretched and contracted to the crossing point, a stub of $e_{2}$ extends to the crossing point, then no intersection will occur.
Let the time it takes for the stub of $e_{1}$ to contract to the crossing point after it starts morphing be $t_{1}$ and the time it takes for the stub of $e_{2}$ to start morphing and then extend to the crossing point be $t_{2}$.
If the stub of edge $e_{2}$ starts morphing $t_{1}-t_{2}$ after the stub of $e_{1}$ starts morphing, no crossing occurs.
In addition, when morphing is repeated alternately, $e_{1}$ will start morphing again next to $e_{2}$.

\subsection{Parallel Morphing}

Even if edges belong to the same morphing group, morphing of two non-intersecting edges may be performed simultaneously.
By appropriately morphing parallelly, it is possible to shorten the total morphing time of all edges.
For example, as edges $e_{1}$ and $e_{2}$ in Fig.~\ref{fig:dependency} are not intersecting, their morphing can be parallelized if we can find adequate timing to avoid crossing with $e_{3}$.
This can reduce the overall time.

\subsection{Algorithm for Finding Morphing Start Time}
\label{ssec:animationStartTime}

In scheduling morphing, we decided to determine the start time from longer edges.
Assuming that every stub has the same morphing speed, the longer the edge, the longer one cycle of morphing takes.
By determining the start time from longer edges, while a long edge is morphing, the morphing of short edges that do not intersect with it can be embedded in the same time range.

Given a morphing group $E$ (a set of edges), Algorithm~\ref{alg:DecideAnimationStartAlg} determines the morphing start time $t_{s}(e)$ for all edges $e\in E$.
The algorithm determines the morphing start time in descending order of edge length.
It checks the timing of every morphing stub of all edges in $C(e)$ that intersect with edge $e$, and allows the morphing start time of edge $e$ to be the earliest time that does not result in intersection with the morphing stubs.

The $r_{1}(e,c)$ and $r_{2}(e,c)$ appearing in the algorithm represents the first and last time of the time range, respectively, when the start of the morphing of edge $e$ is prohibited to avoid crossing with edge $c$.
They are described as Exp.~(\ref{exp:rs}) and Exp.~(\ref{exp:re}).

\begin{align}
  r_{1}(e,c) &= t_{s}(c) + t_{p}(c,e) - t_{r}(e,c) \label{exp:rs} \\
  r_{2}(e,c) &= t_{s}(c) + t_{r}(c,e) - t_{p}(e,c) \label{exp:re},
\end{align}
where $t_{p}(e,c)$ is the time it takes from the start of morphing of the stub of edge $e$ to the first passing (passing while stretching) at the crossing point with edge $c$ (cf. $t_{p}(e_{2},e_{1})=t_{2}$ in Fig.~\ref{fig:just_non_crossing}), and
$t_{r}(e,c)$ is the time it takes from the start of morphing to the second passing (passing while shrinking) at the crossing point (cf. $t_{r}(e_{1},e_{2})=t_{1}$ in Fig.~\ref{fig:just_non_crossing}).

\begin{algorithm}[htb]
  \caption{Determining the start time of morphing}
  \label{alg:DecideAnimationStartAlg}
  \begin{algorithmic}[1]
    \Input{$E$ --- Set of edges included in a morphing group}
    \Output{Start time is determined for all edges of $E$}
    \Function{findStartTime}{$E$}
      \For{$e$ in $sortByLength(E)$}
        \State $I \leftarrow \{(r_{1}(e,c), r_{2}(e,c))|c\in C(e)\wedge t_{s}(c)$ is defined.$\}$       
        \State $t_{s}(e) \leftarrow earliestSpace(I)$
      \EndFor
    \EndFunction
  \end{algorithmic}
\end{algorithm}

Function $earliestSpace(I)$ yields the smallest value not included in the time ranges (intervals) in a given set $I$.
If each pair $(r_{1},r_{2})$ included in the set $I$ is regarded as an interval $[r_{1},r_{2})$ of real numbers, function $earliestSpace(I)$ is defined as Exp.~(\ref{exp:earliestSpace}).
We calculate $earliestSpace(I)$ using Algorithm~\ref{alg:earliestSpaceAlg}.
Note that $T$ is a nonnegative real number in Algorithm~\ref{alg:earliestSpaceAlg}.

\begin{equation}
  earliestSpace(I) = \min\left\{\left(\bigcup_{r\in I} r \right) ^{c}\right\} \label{exp:earliestSpace}
\end{equation}

\begin{algorithm}[htb]
  \caption{Finding time when morphing can start}
  \label{alg:earliestSpaceAlg}
  \begin{algorithmic}[1]
    \Input{$I$ --- Set of time ranges (pairs of times) during which morphing should not start}
    \Output{ Earliest time morphing can start}
    \Function{earliestSpace}{$I$}
      \State $t \leftarrow 0$
      \For{$(r_{1}, r_{2})$ in $sortByStartTime(I)$}
        \If{$r_{2} < t$}
          \State continue
        \ElsIf{$t < r_{1}$}
          \State \Return $t$
        \Else
          \State $t \leftarrow r_{2}$
        \EndIf
      \EndFor
      \State \Return $t$
    \EndFunction
  \end{algorithmic}
\end{algorithm}

\section{Evaluation Experiment}

To investigate the effectiveness of MED, we conducted a comparative experiment with three types of visual representations: CED, 1/4-SHPED, and 1/4-SHMED.

\subsection{Hypothesis}
We made the following hypothesis.

\begin{description}
  \item[H1] 1/4-SHMED requires less time to read a graph than 1/4-SHPED
  \item[H2] 1/4-SHMED is more accurate at reading graphs than CED
\end{description}

\subsection{Tasks}\label{sec:task}

We designed the following tasks to test the above hypotheses.

\begin{description}
\item[T1] Check if the two highlighted nodes are adjacent (connected by an edge).
\item[T2] Select all the nodes to which the highlighted node is adjacent.
\end{description}
  
\begin{figure}[htb]
\centering
\begin{tabular}{ccc}
\includegraphics[width=0.33\linewidth]{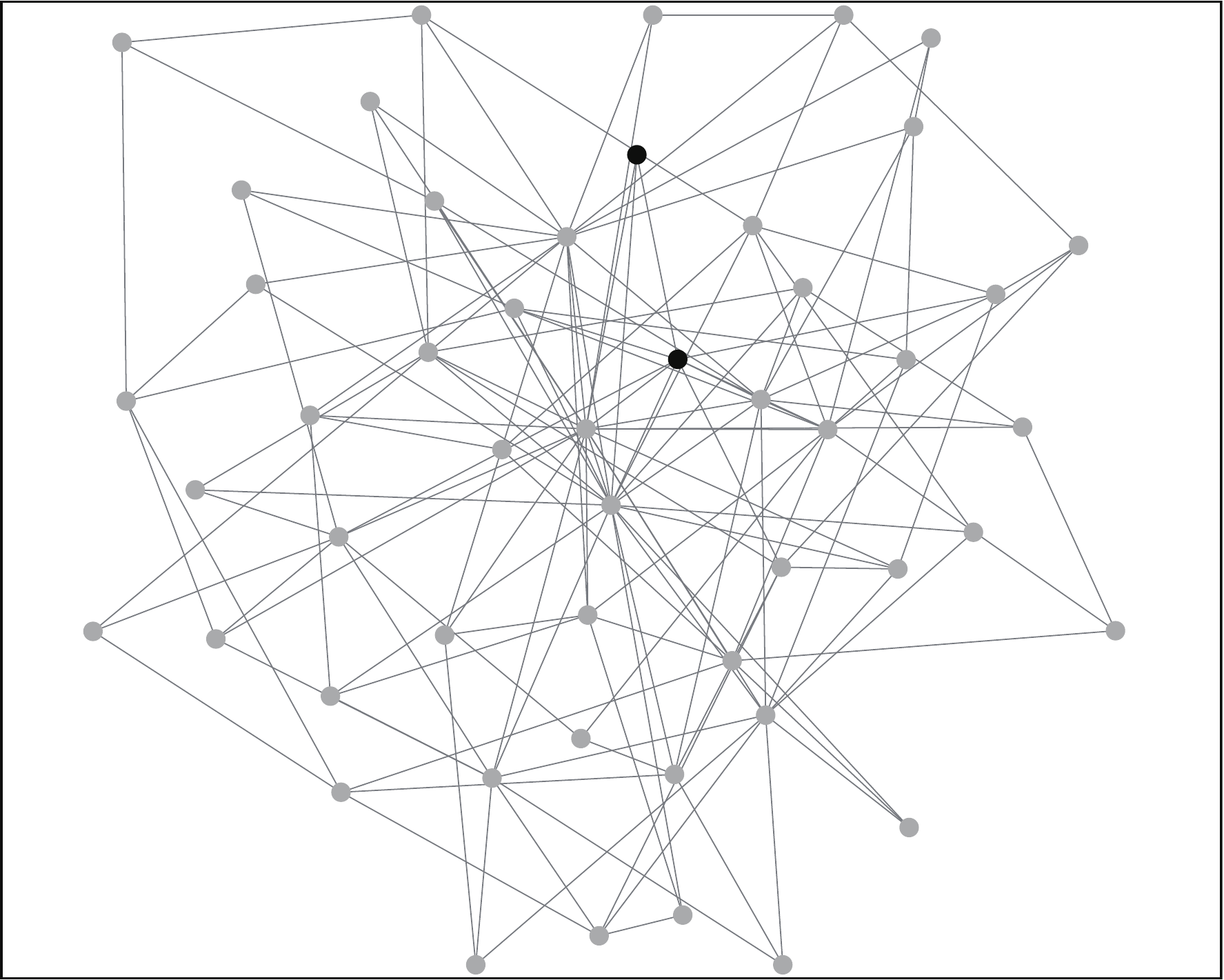} &
\includegraphics[width=0.33\linewidth]{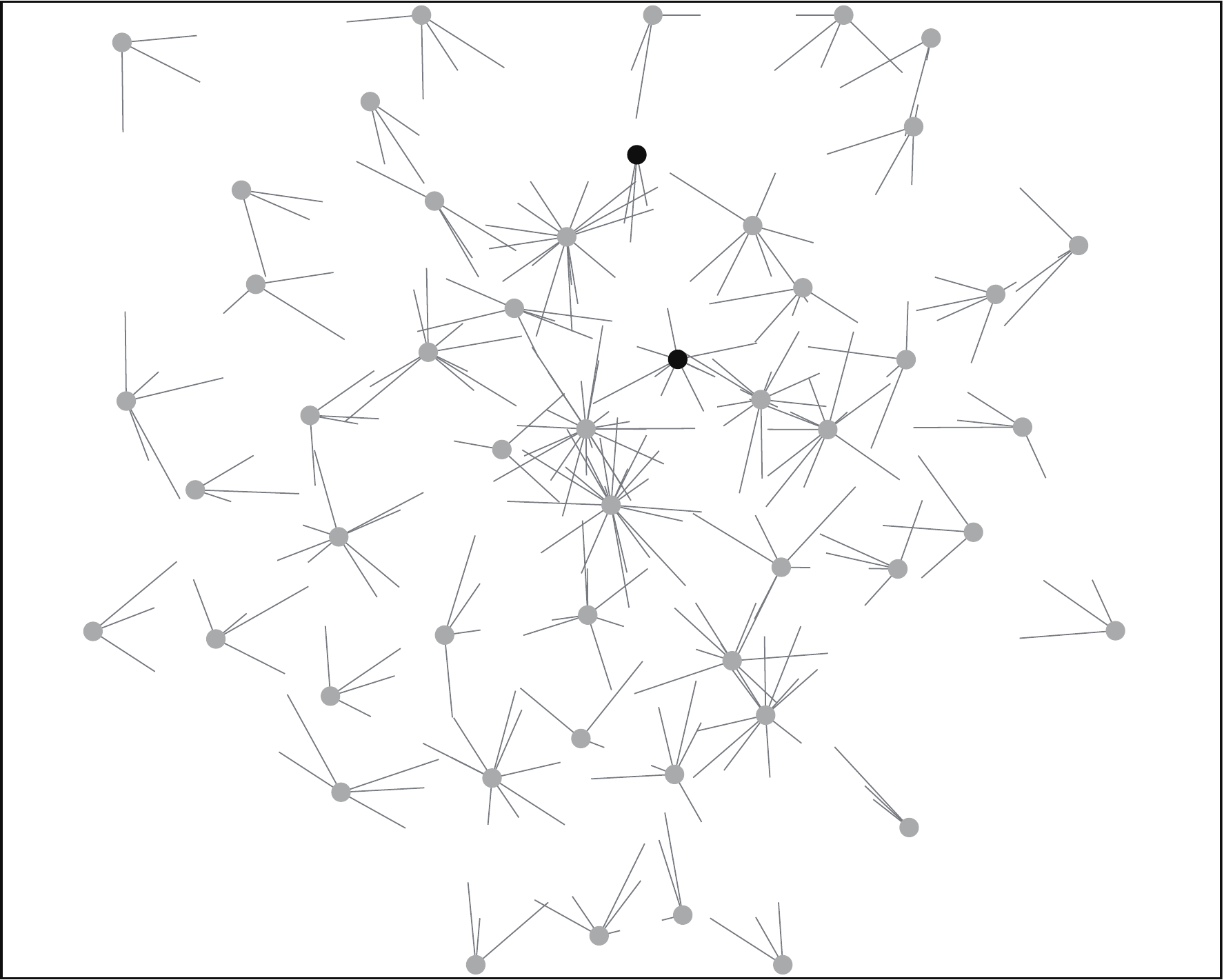} &
\includegraphics[width=0.33\linewidth]{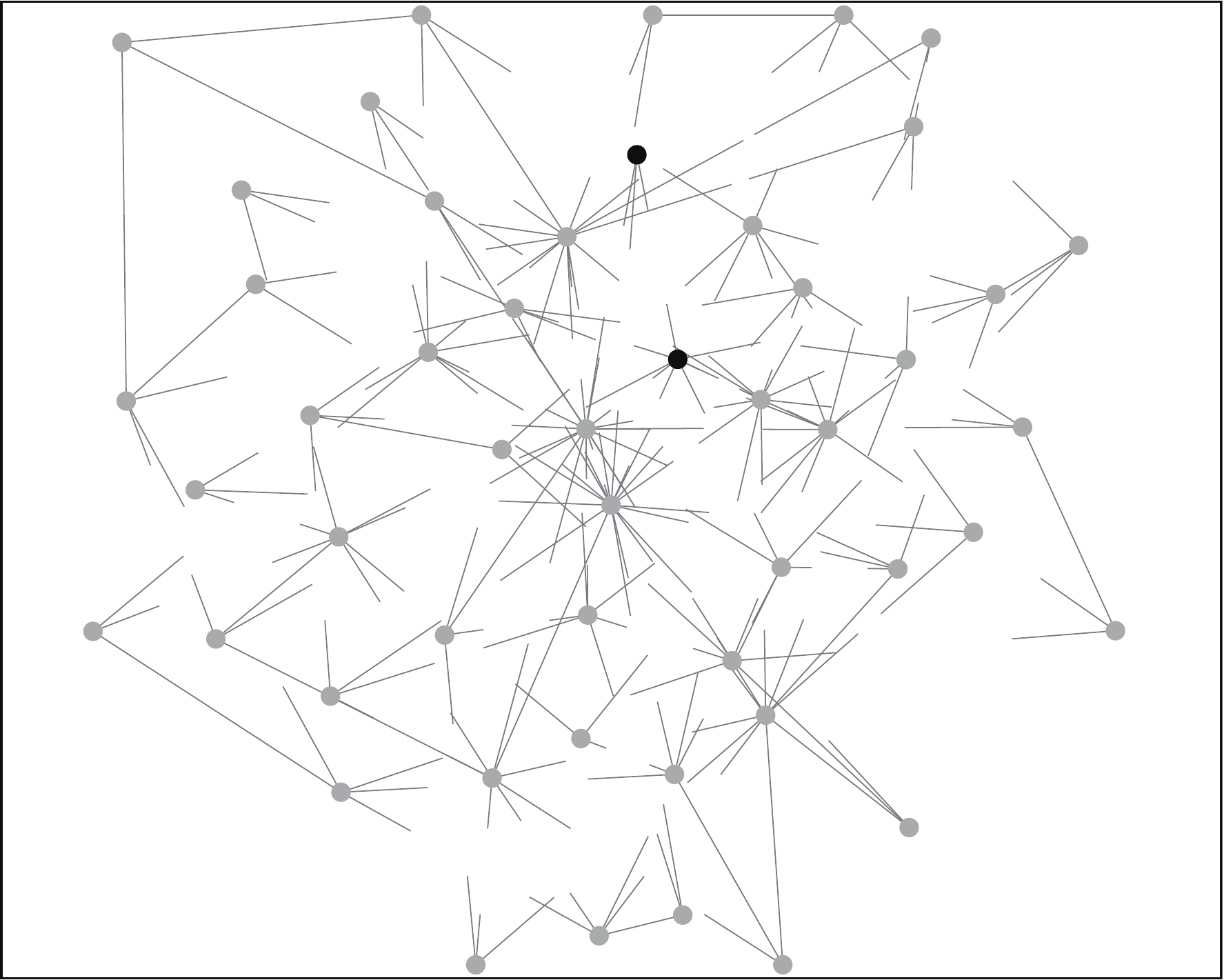} \\
(a) CED & (b) 1/4-SHPED & (c) 1/4-SHMED \\
\end{tabular}
\caption{Examples of visual representations used in T1}
\label{fig:Task1Img}
\bigskip
 
\begin{tabular}{ccc}
\includegraphics[width=0.33\linewidth]{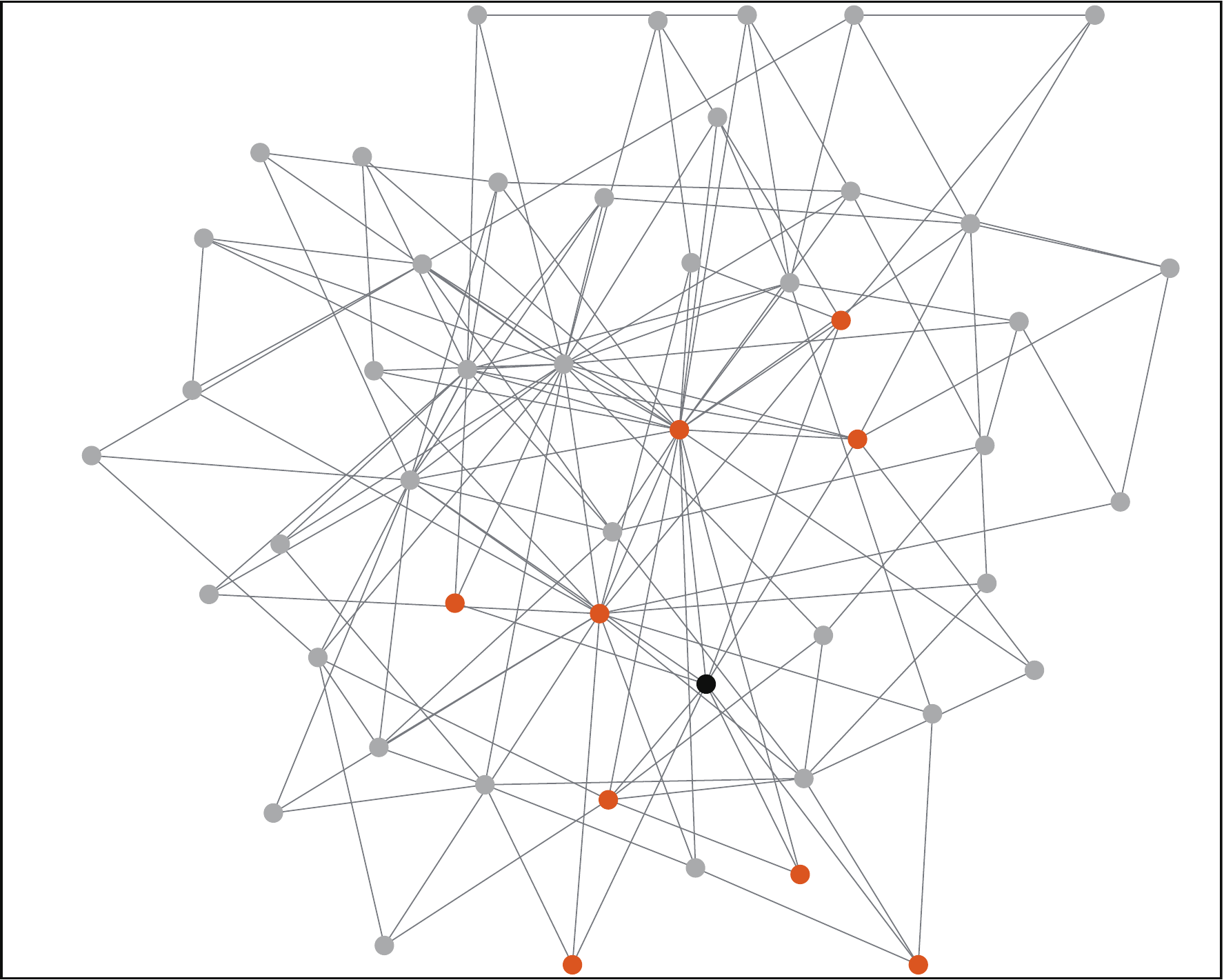} & 
\includegraphics[width=0.33\linewidth]{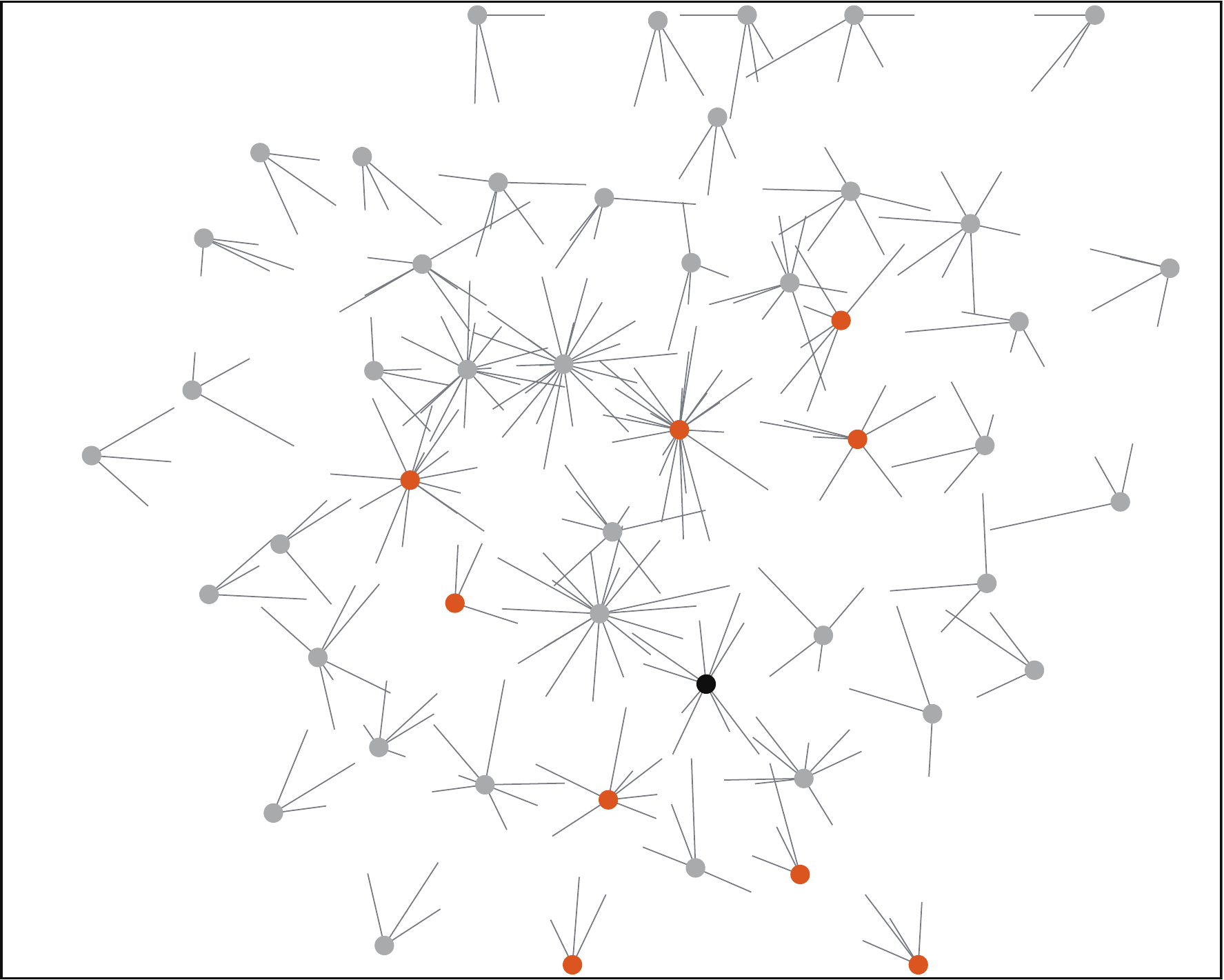} &
\includegraphics[width=0.33\linewidth]{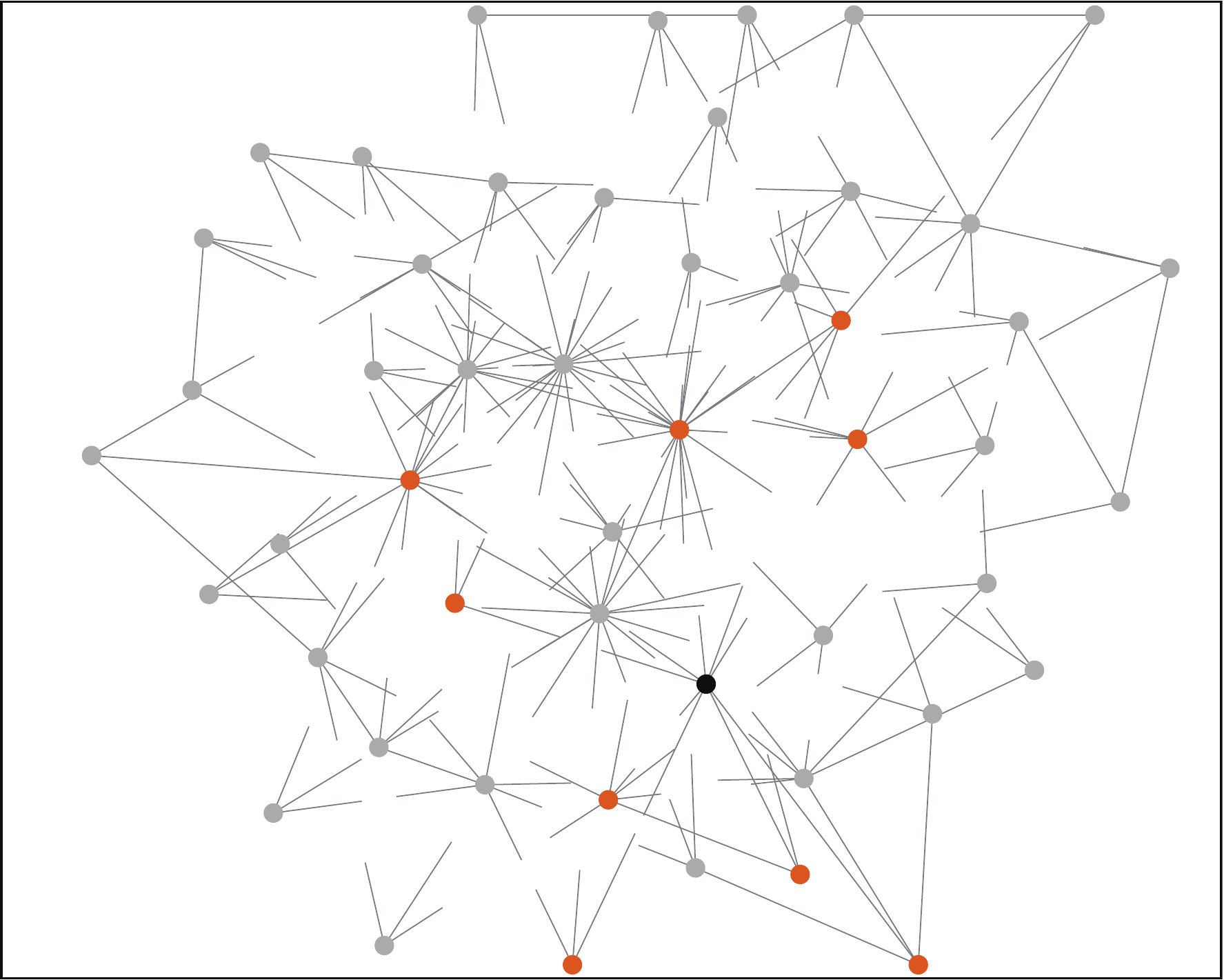} \\
(a) CED & (b) 1/4-SHPED & (c) 1/4-SHMED \\
\end{tabular}
\caption{Examples of visual representations used in T2}
\label{fig:Task2Img}

\end{figure}

For T1, as shown in Fig.~\ref{fig:Task1Img}, a graph in which two nodes are highlighted is displayed.
Participants respond by pressing ``Y'' or ``N'' on the keyboard.
When creating sample graphs, the number of crossings of the edges connecting two nodes were set to 8 or 16 when the nodes were adjacent.

For T2, as shown in Fig.~\ref{fig:Task2Img}, a graph in which one node is highlighted is displayed.
Node selection is performed using a trackpad.
When clicked, the pointed node is selected and turns orange.
Participants can also cancel the selection by clicking again.
Answers are confirmed by pressing the Enter key.
When creating sample graphs, we selected nodes to be highlighted such that the number of adjacent nodes to it were 3, 6, and 9.
Furthermore, we set the average number of intersections of the edges of interest to be within 7.9--8.1 and the average of lengths of the edges to be in the range of 3.3--3.7~cm on the screen to ensure that the task difficulty was not excessively low or high.
In this experiment, we assumed that the distance between the participant's eyes and the screen was 40~cm.

\subsection{Graphs used for the experiment}\label{ssec:MakingGraph}

We used the Barab{\'a}si-Albert model~\cite{barabasi1999emergence} as a guideline to create a graph with 50 nodes and 144 edges.
We used the Fruchterman-Reingold algorithm~\cite{fruchterman1991graph} to determine the layout of the graph.

\subsection{Morphing speed}\label{ssec:AnimSpeed}

We set the morphing speed of each stub as $10^\circ$/s.
If morphing is too fast, the human eye cannot track it.
Conversely, if it is too slow, reading efficiency is reduced.
We derived the speed based on Robinson's experiment~\cite{robinson1965mechanics} such that it is human eye-trackable while being as fast as possible.
However, we set a minimum one-way travel time 300 ms to make capturing morphing stubs easy.

\subsection{Experimental settings}
We used a MacBook Pro 2017 (screen size 13.3 inches, screen resolution $1440\times 900$) for the experiment.
We set the display area to $1000\times 800$ so the graph can be viewed without scrolling.

The participants in this experiment were 12 students (4 university students and 8 graduate students).

\subsection{Experimental procedure}
The following procedure was used to conduct the experiment:

\begin{description}
\item[1] Overall explanation
\item[2] Visual representation \#1
  \begin{description}
  \item[2-1] T1 practice (one question)
  \item[2-2] T1 actual (nine questions)
  \item[2-3] T2 practice (one question)
  \item[2-4] T2 actual (nine questions)
  \item[2-5] Questionnaire for visual representation \#1
  \end{description}
\item[3] Visual representation \#2
  (flow similar to visual representation \#1)
\item[4] Visual representation \#3
  (flow similar to visual representation \#1)
\item[5] Questionnaire for whole experiment
\end{description}

We varied the order of presenting visual representations from each participant to eliminate the effects of order.
Therefore, visual representations \#1, \#2, and \#3 differ depending on the participant.
We assigned two participants for each of the six ($=3!$) orders.

\subsection{Response time}

\begin{figure}[htb]
  \centering
  \begin{tabular}{cc}
  \includegraphics[scale=0.3]{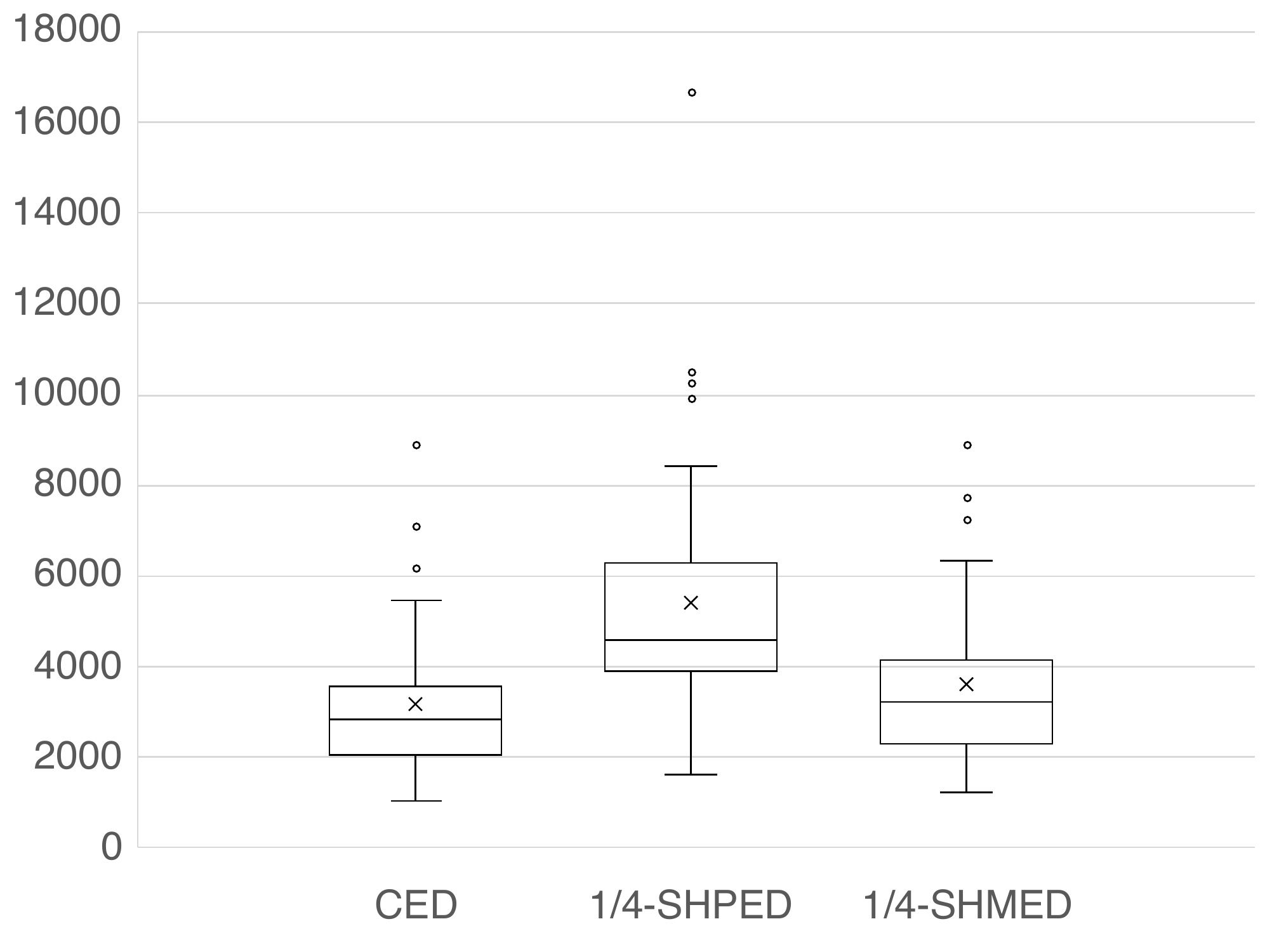} &
  \includegraphics[scale=0.3]{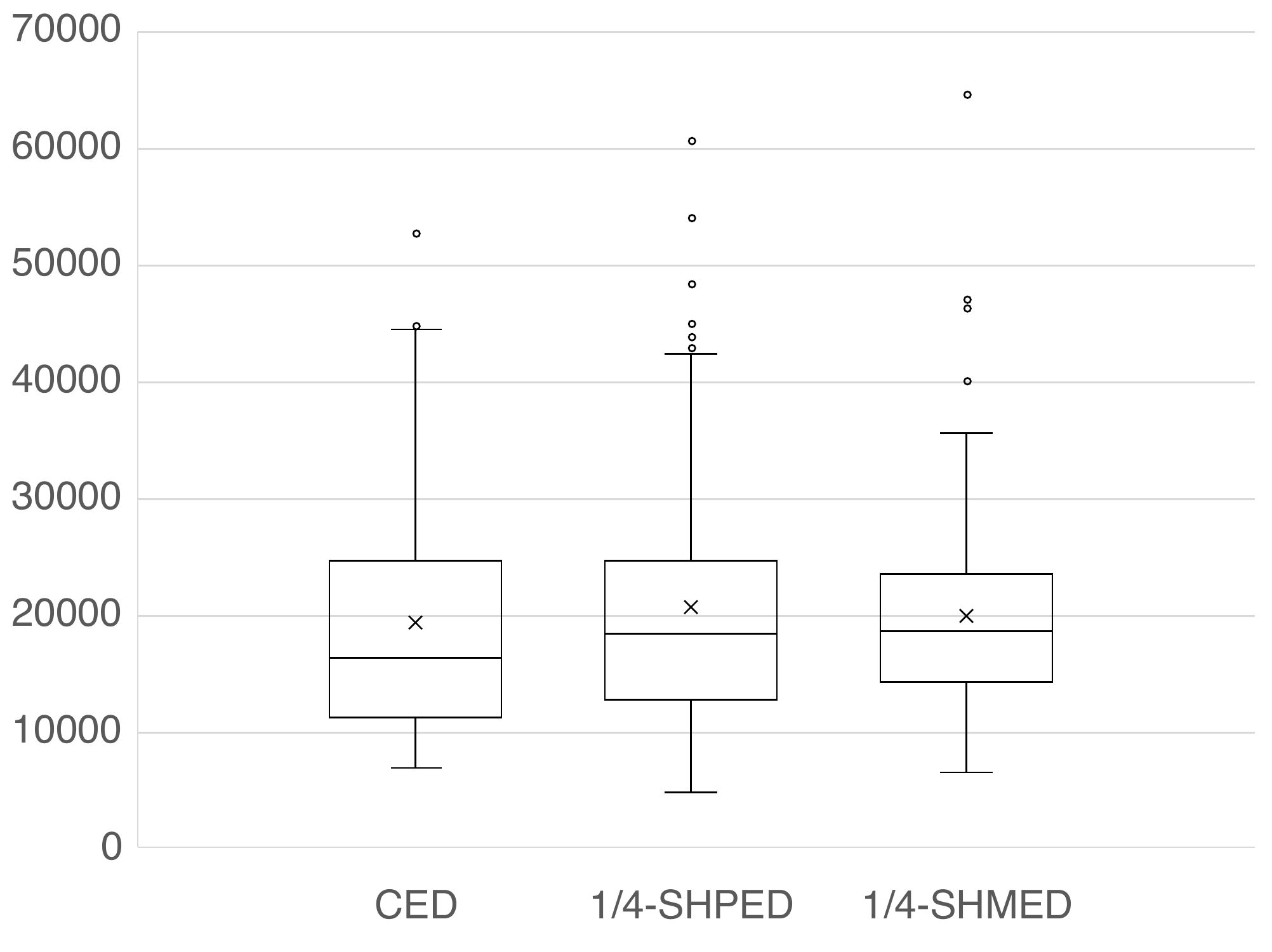} \\
  (a) Task T1 [ms] & (b) Task T2 [ms] \\
  \end{tabular}
  \caption{Distribution of response time}
  \label{fig:task_time}
\end{figure}

Fig.~\ref{fig:task_time} shows the distribution (boxplots) of response time (in millisecond) for each task and representation method.
In both tasks, the average response time was the lowest for CED and highest for 1/4-SHPED.
As the 1/4-SHMED is located in the middle, an improvement in the reading time for 1/4-SHPED can be expected.
From the Shapiro-Wilk test ($\alpha =0.05$), the time taken either task did not follow a normal distribution.
Therefore, we performed multiple tests using the Friedman and Holm methods.
Tables~\ref{tbl:T1TimeTest} and \ref{tbl:T2TimeTest} show the test results for the response time for T1 and T2, respectively.
As shown in Table~\ref{tbl:T1TimeTest}, a significant difference was observed between the representation methods,
i.e., 1/4-SHMED can shorten the time taken to confirm the adjacency between nodes, compared to 1/4-SHPED (H1).
In contrast, no significant difference was found between the representation methods with respect to the response time of T2.

\begin{table}[htb]
  \centering
    \caption{Test result of response time of T1}
    \label{tbl:T1TimeTest}
    \begin{tabular}{r@{ vs }l|cc} \hline
      \multicolumn{2}{c|}{Comparison} & Test result (p value) & Significance level \\ \hline \hline
      CED & 1/4-SHPED       & 2.035e-7 & $<$ 0.0167 \\ \hline
      CED & 1/4-SHMED       & 0.0343 & $<$ 0.0500 \\ \hline
      1/4-SHPED & 1/4-SHMED & 0.0011 & $<$ 0.0250 \\ \hline
    \end{tabular}
\bigskip
    
    \caption{Test result of response time of Task T2}
    \label{tbl:T2TimeTest}
    \begin{tabular}{r@{ vs }l|cc} \hline
      \multicolumn{2}{c|}{Comparison} & Test result (p value) & Significance level \\ \hline \hline
      CED & 1/4-SHPED       & 0.0543 & $>$ 0.0167 \\ \hline
      CED & 1/4-SHMED       & 0.1237 & --- \\ \hline
      1/4-SHPED & 1/4-SHMED & 0.8474 & --- \\ \hline
    \end{tabular}
\end{table}

\subsection{Answer accuracy}
\begin{figure}[htb]
  \centering
  \begin{tabular}{cc}
  \includegraphics[scale=0.3]{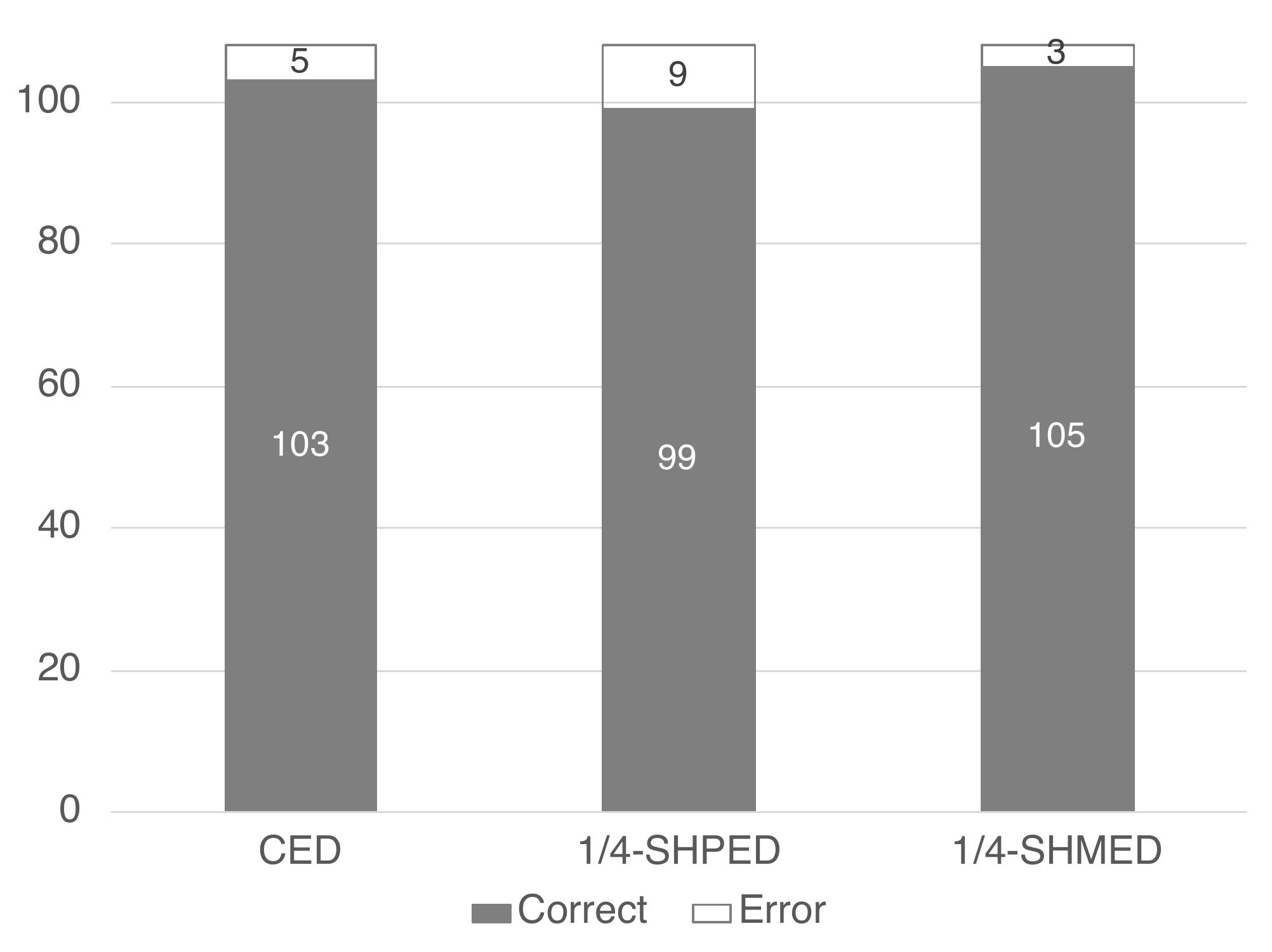} &
  \includegraphics[scale=0.3]{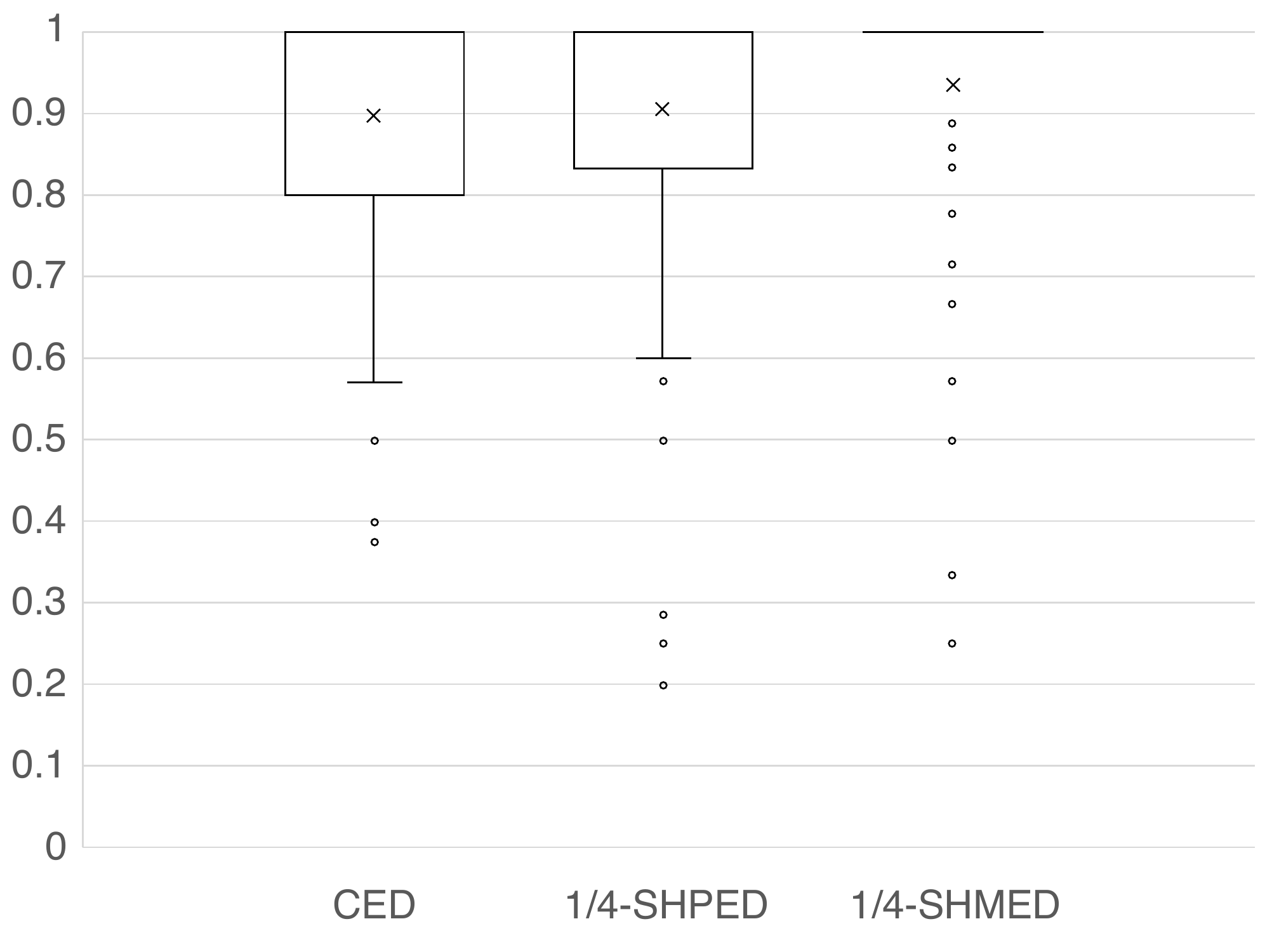} \\
  (a) Task T1 & (b) Task T2 \\
  \end{tabular}
  \caption{Correct answer rate}
  \label{fig:task_score}
\end{figure}

Fig.~\ref{fig:task_score}(a) shows the number of correct answers and the number of incorrect answers for T1 in a stacked bar chart.
The correct answer rate of 1/4-SHMED is the highest.
However, independence between the representation methods was not recognized from the chi-square test.
We defined the score for T2 as the Jaccard coefficient between the set of adjacent nodes and the set of answered nodes,
i.e., it is 1 when the two sets completely match, and 0 when there is no common element.
Fig.~\ref{fig:task_score}(b) shows the distribution of scores according to each representation method for T2 in a boxplot.
From the Shapiro-Wilk test ($\alpha =0.05$), T2 did not followed a normal distribution.
Therefore, we performed multiple tests using the Friedman and Holm methods.
Table~\ref{tbl:task2score_test} shows the test results for the scores T2.
As seen above, regarding the accuracy of answers, no significant difference was observed between the representation methods.
Therefore, H2 is not supported.

\begin{table}[htb]
  \centering
    \caption{Test result of score of T2}
    \label{tbl:task2score_test}
    \begin{tabular}{r@{ vs }l|cc} \hline
      \multicolumn{2}{c|}{Comparison} & Test result (p value) & Significance level \\ \hline \hline
            CED & 1/4-SHPED & 0.5862 & ---  \\ \hline
            CED & 1/4-SHMED & 0.1489 & ---  \\ \hline
      1/4-SHPED & 1/4-SHMED & 0.07817 & $>$ 0.0167 \\ \hline
    \end{tabular}
\end{table}

\subsection{Qualitative Feedback}

We asked the participants for opinions on visual representations using questionnaires.
The following comments were obtained on 1/4-SHMED.

\begin{itemize}
\item Positive opinions
\begin{itemize}
\item Morphing made it easy to confirm the exact adjacency.
\item It can be judged whether two nodes are adjacent by observing the morphing of two stabs works simultaneously.
\end{itemize}

\item Negative opinions
\begin{itemize}
\item It is messy and difficult to see. My eyes are strained. 
\item The stubs change too fast. The time for stubs to connect is too short.
\end{itemize}

\end{itemize}

Positive opinions indicate that morphing contributes to reading graphs.
In contrast, from the negative opinions, it appears that visual clutter was not always resolved.
The following can be considered as the main reasons.
The first is that the morphing speed is too fast.
In the implementation used for the experiment, to shorten the overall morphing time, the morphing speed was determined based on the tracking speed of human eyes; however, this appears to be too fast.
The second is that there were a large number of stubs applying morphing.
In the graph used in the experiment, out of the 144 edges, the average number of non-morphing edges is 24.5.
Given that approximately 120 edges repeated morphing, the entire graph is considered to have caused visual clutter.

\section{Concluding Remarks}

We proposed morphing edge drawing (MED) which is time-varying partial edge drawing (PED) and showed the formalization of MED.
We also developed a scheduling scheme for morphing such that dynamic stubs do not cause new crossings.
We compared three visual representations, CED, 1/4-SHPED, and 1/4-SHMED, via a user study, and showed that 1/4-SHMED is better than 1/4-SHPED in terms of graph reading time.
Thus, MED can function as a countermeasure against the time to read a graph by PED.
In the future, it is important to investigate eye-friendly morphing that causes less strain and has improved scheduling.

%
%
%
 \bibliographystyle{splncs04}
 \bibliography{reference}

\begin{thebibliography}{10}
\providecommand{\url}[1]{\texttt{#1}}
\providecommand{\urlprefix}{URL }
\providecommand{\doi}[1]{https://doi.org/#1}

\bibitem{barabasi1999emergence}
Barab{\'a}si, A.L., Albert, R.: Emergence of scaling in random networks.
  Science  \textbf{286}(5439),  509--512 (1999).
  \doi{10.1126/science.286.5439.509}

\bibitem{becker1995visualizing}
{Becker}, R.A., {Eick}, S.G., {Wilks}, A.R.: Visualizing network data. IEEE
  Transactions on Visualization and Computer Graphics  \textbf{1}(1),  16--28
  (March 1995). \doi{10.1109/2945.468391}

\bibitem{binucci2016partial}
{Binucci}, C., {Liotta}, G., {Montecchiani}, F., {Tappini}, A.: Partial edge
  drawing: Homogeneity is more important than crossings and ink. In: 2016 7th
  International Conference on Information, Intelligence, Systems Applications
  (IISA). pp.~1--6 (July 2016). \doi{10.1109/IISA.2016.7785427}

\bibitem{blass2009smooth}
{Blaas}, J., {Botha}, C., {Grundy}, E., {Jones}, M., {Laramee}, R., {Post}, F.:
  Smooth graphs for visual exploration of higher-order state transitions. IEEE
  Transactions on Visualization and Computer Graphics  \textbf{15}(6),
  969--976 (Nov 2009). \doi{10.1109/TVCG.2009.181}

\bibitem{bruckdorfer2012mad}
Bruckdorfer, T., Kaufmann, M.: Mad at edge crossings? break the edges! In:
  Kranakis, E., Krizanc, D., Luccio, F. (eds.) Fun with Algorithms. pp. 40--50.
  Springer Berlin Heidelberg, Berlin, Heidelberg (2012)

\bibitem{bruckdorfer2015ped}
Bruckdorfer, T., Kaufmann, M., Leib{\ss}le, S.: {PED} user study. In:
  Di~Giacomo, E., Lubiw, A. (eds.) Graph Drawing and Network Visualization. pp.
  551--553. Springer International Publishing, Cham (2015)

\bibitem{burch2017user}
{Burch}, M.: A user study on judging the target node in partial link drawings.
  In: 2017 21st International Conference Information Visualisation (iV). pp.
  199--204 (July 2017). \doi{10.1109/iV.2017.43}

\bibitem{burch2012evaluating}
Burch, M., Vehlow, C., Konevtsova, N., Weiskopf, D.: Evaluating partially drawn
  links for directed graph edges. In: van Kreveld, M., Speckmann, B. (eds.)
  Graph Drawing. pp. 226--237. Springer Berlin Heidelberg, Berlin, Heidelberg
  (2012)

\bibitem{collins2009parallel}
Collins, C., Vi\'{e}gas, F.B., Wattenberg, M.: Parallel tag clouds to explore
  and analyze faceted text corpora. In: 2009 IEEE Symposium on Visual Analytics
  Science and Technology. pp. 91--98 (Oct 2009).
  \doi{10.1109/VAST.2009.5333443}

\bibitem{fruchterman1991graph}
Fruchterman, T.M.J., Reingold, E.M.: Graph drawing by force-directed placement.
  Software: Practice and Experience  \textbf{21}(11),  1129--1164 (1991).
  \doi{10.1002/spe.4380211102}

\bibitem{holten2011extended}
{Holten}, D., {Isenberg}, P., {van Wijk}, J.J., {Fekete}, J.: An extended
  evaluation of the readability of tapered, animated, and textured
  directed-edge representations in node-link graphs. In: 2011 IEEE Pacific
  Visualization Symposium. pp. 195--202 (March 2011).
  \doi{10.1109/PACIFICVIS.2011.5742390}

\bibitem{robinson1965mechanics}
Robinson, D.A.: The mechanics of human smooth pursuit eye movement. The Journal
  of Physiology  \textbf{180}(3),  569--591 (1965).
  \doi{10.1113/jphysiol.1965.sp007718}

\bibitem{Romat:2018:AET:3173574.3173761}
Romat, H., Appert, C., Bach, B., Henry-Riche, N., Pietriga, E.: Animated edge
  textures in node-link diagrams: A design space and initial evaluation. In:
  Proceedings of the 2018 CHI Conference on Human Factors in Computing Systems.
  pp. 187:1--187:13. CHI '18, ACM, New York, NY, USA (2018).
  \doi{10.1145/3173574.3173761},
  \url{http://doi.acm.org/10.1145/3173574.3173761}

\bibitem{schmauder2015visualizing}
Schmauder, H., Burch, M., Weiskopf, D.: Visualizing dynamic weighted digraphs
  with partial links. In: Proceedings of 6th International Conference on
  Information Visualization Theory and Applications (IVAPP). pp. 123--130
  (2015)

\end{thebibliography}

\end{document}